\documentclass[aps,pra,showpacs,twoside,twocolumn,longbibliography,10pt]{revtex4-2}

\usepackage[colorlinks=true,citecolor=blue,urlcolor=blue,linkcolor=blue]{hyperref}

\usepackage{epsfig,newlfont,amssymb,amsfonts,amsmath,bm,subfigure,palatino,mathtools,amsthm,braket,times,soul,enumitem,color}

\usepackage[normalem]{ulem}
\usepackage[english]{babel}
\usepackage[utf8]{inputenc}

\usepackage{array}
\usepackage{xcolor}
\usepackage{graphics}
\usepackage{bbm}
\usepackage{wrapfig}
\usepackage{hyphenat}
\usepackage{float}
\usepackage{orcidlink}
\usepackage{verbatim}
\usepackage{natbib}

\def\Tr{\mathrm{Tr}}

\begin{document}

\title{Generating Superposition of Transverse Ising Ground States by Quantum Interaction}

\author{Koushik Das}
\author{Sayan Mondal}
\author{Ujjwal Sen}

\affiliation{Harish-Chandra Research Institute, Chhatnag Road, Jhunsi, Prayagraj  211 019, India\\
Homi Bhabha National Institute, Training School Complex, Anushakti Nagar, Mumbai
400 094, India}

\begin{abstract}
{We investigate superpositions of quantum Ising ground states generated through a \emph{quantum interaction}, in which the nearest-neighbor coupling strength is promoted to an operator acting on a central auxiliary spin. We prepare the central spin in a superposition, which allows the chain to evolve coherently under different effective interaction strengths. The states are prepared by adiabatic evolution of the joint system comprising the central spin and the many-body spin chain, followed by a measurement on the central spin and post-selection of a specific outcome. It is possible to prepare superpositions of transverse-field quantum Ising ground states belonging to the three magnetic phases, namely, the paramagnetic, ferromagnetic, and anti-ferromagnetic phases. For a spin-$1/2$ central system, it is possible to prepare superpositions of ground states belonging to any two phases, while for the spin-$1$ case, the superpositions may contain three ground states, each belonging to one of the three phases. We analytically derive expressions for the magnetization, spin correlations, and entanglement of such states. The magnetization and spin correlations show distinct features in different superposition sectors. The nearest-neighbor entanglement is largest where the transverse field and Ising interaction compete, and its behavior differs from that of the magnetic observables. We also derive the quantum Fisher information relevant to estimating the Ising interaction strength. And it exhibits multiple peaks near the critical points of the constituent ground states, with the maximum value showing Heisenberg scaling, for the parameters and system sizes examined.
}
\end{abstract}

\maketitle

\section{Introduction}
In quantum many-body systems, a quantum phase transition~\cite{Carr2010,Sachdev2011,Dutta2015} can occur at zero temperature when a parameter of the Hamiltonian is varied across a critical point in the thermodynamic limit. The one-dimensional transverse-field Ising model is a simple many-body system with nearest-neighbor interactions and a local transverse field acting on each spin. The competition between the interaction and the field gives rise to ordered and disordered phases. Depending on the sign of the interaction coupling, the ordered phase can be either ferromagnetic or anti-ferromagnetic, while the disordered phase is paramagnetic. The model has been extensively studied for its quantum critical behavior, entanglement properties, and non-equilibrium dynamics~\cite{Osterloh2002,Vidal2003,Dziarmaga2005,Quan2006,Calabrese2011,Rams2011,Heyl2013}. It has also been investigated in quantum battery charging~\cite{Mondal2022,Mitra2024}, quantum sensing~\cite{Invernizzi2008,Mishra2022}, and quantum thermal machines~\cite{Revathy2020,Piccitto2022}, and has served as a benchmark for quantum annealing and variational algorithms~\cite{Kadowaki1998,Ho2019,Sumeet2024}.

In this regard, an interesting question is whether ground states belonging to different magnetic phases can be prepared in a coherent superposition, and how the properties of such a state depend on its components. These states allow different forms of magnetic order to contribute coherently, with interference terms entering the expectation values of observables. It is therefore useful to investigate how this coherence affects their magnetization, spin correlations, and entanglement. Their sensitivity to changes in the Hamiltonian parameters also raises the possibility of using such states for quantum sensing. Addressing these questions requires a preparation scheme that allows control over both the constituent ground states and their relative amplitudes.

\begin{figure}[ht]
    \centering
    \includegraphics[width=0.4\textwidth]{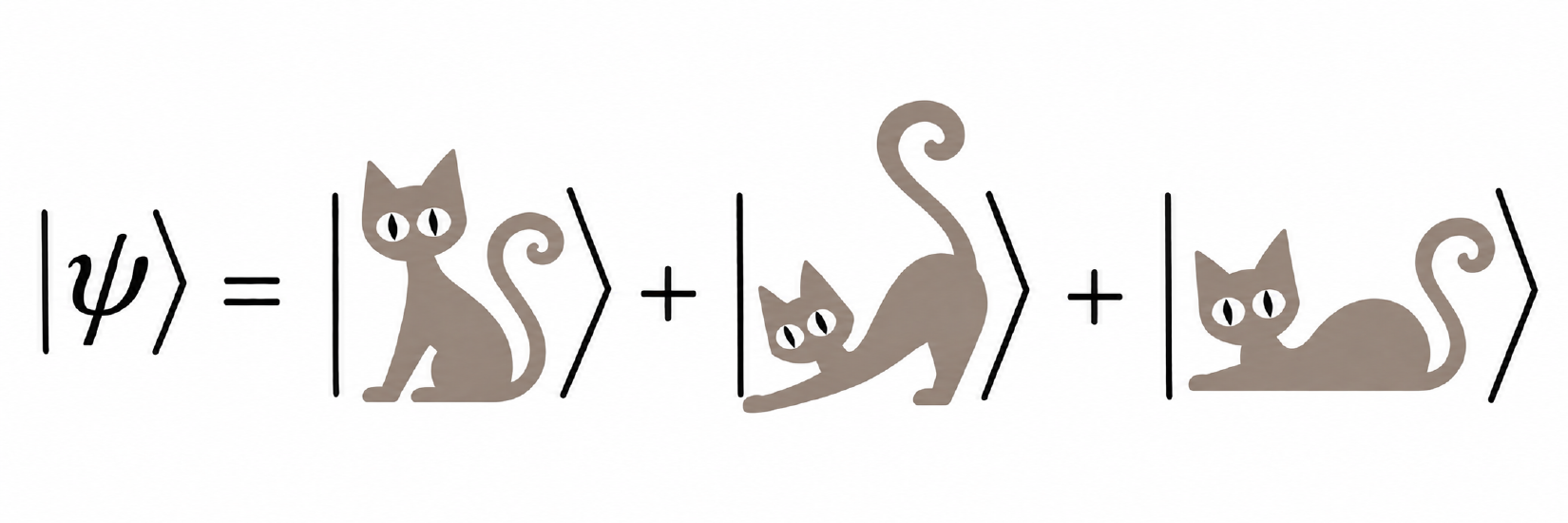}
    \caption{A cat can be sitting, stretching, or lying down, representing three different \emph{phases} in this illustration. Its behavior differs between these phases, but how would it behave in a superposition of them? In this work, we investigate coherent superpositions of ground states of the transverse-field Ising model belonging to the ferromagnetic, anti-ferromagnetic, and paramagnetic phases, and study the physical properties of the resulting states.}
    \label{fig:cat-superposition}
\end{figure}

A common approach for preparing many-body ground states is through adiabatic evolution~\cite{Schaller2008,Lubasch2011,Farooq2015,Albash2018}, in which the Hamiltonian is varied sufficiently slowly to allow the system to follow its instantaneous ground state. In conventional adiabatic protocols, the control parameters are treated as classical quantities with definite values at each instant. In this work, we introduce a \emph{quantum interaction} by promoting the Ising coupling strength to an operator acting on a central spin. When the central spin is initially prepared in a superposition, adiabatic evolution correlates each central-spin component with the ground state of the chain at a different effective interaction strength. A suitable projective measurement of the central spin, followed by post-selection of a specific outcome, then leaves the chain in a coherent superposition of these ground states. Depending on the effective interaction strengths, the components can belong to different magnetic phases. 

Magnetization and spin correlations provide information about the magnetic properties of quantum many-body states~\cite{Pfeuty1970,Barouch1971}. We investigate the magnetization and nearest-neighbor spin correlations of the prepared states across the different superposition sectors to understand how the magnetic properties of the constituent ground states appear in the superposition. We also examine nearest-neighbor concurrence~\cite{Osborne2002,Osterloh2002} and the generalized geometric measure~\cite{SenDe2010} to characterize bipartite and genuine multipartite entanglement~\cite{Amico2008, DeChiara2018}, respectively. Finally, we evaluate the quantum Fisher information~\cite{Montenegro_2025, Agarwal_2025, ghosh2026} with respect to the Ising interaction strength to study how superposing ground states at different parameter values affects the sensitivity to this coupling.
We find that the different superposition sectors exhibit distinct magnetic and entanglement properties. The nearest-neighbor concurrence is largest where the transverse field and Ising interaction compete.
The quantum Fisher information exhibits additional peaks near the critical points of the constituent ground states. For the parameters and system sizes examined, its maximum value shows scaling consistent with quadratic growth in the chain length, indicating Heisenberg scaling for the quantum Fisher information of the post-selected states. These results point to the possibility of preparing and controlling many-body states with potential uses in quantum information and sensing tasks.

\begin{figure*}[t]
\centering

\begin{minipage}{0.38\textwidth}
    \centering
    \begin{tikzpicture}
        \node[inner sep=0, outer sep=0] (img)
            {\includegraphics[width=0.8\linewidth]{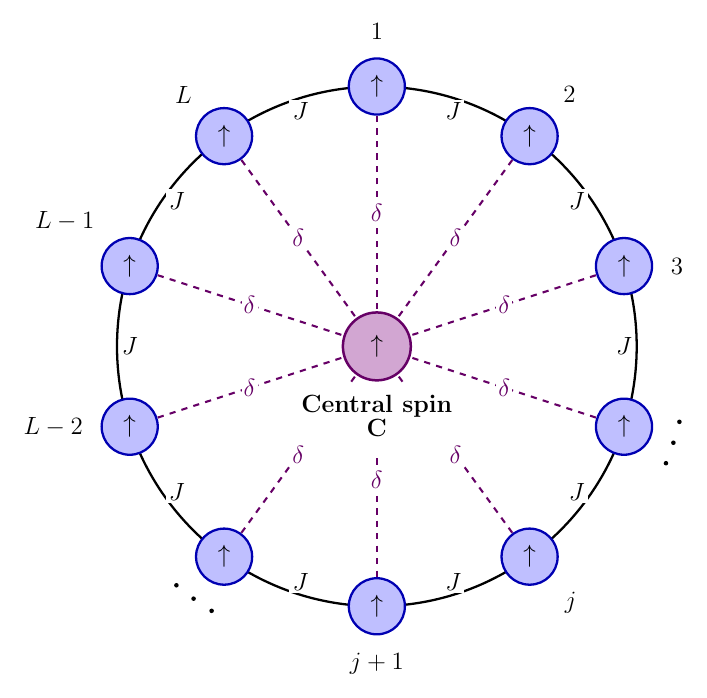}};

        \node[anchor=north west, font=\bfseries]
            at ([xshift=2mm,yshift=-2mm]img.north west) {(a)};
    \end{tikzpicture}
\end{minipage}
\hfill
\begin{minipage}{0.6\textwidth}
    \centering
    \begin{tikzpicture}
        \node[inner sep=0, outer sep=0] (img)
            {\includegraphics[width=\linewidth]{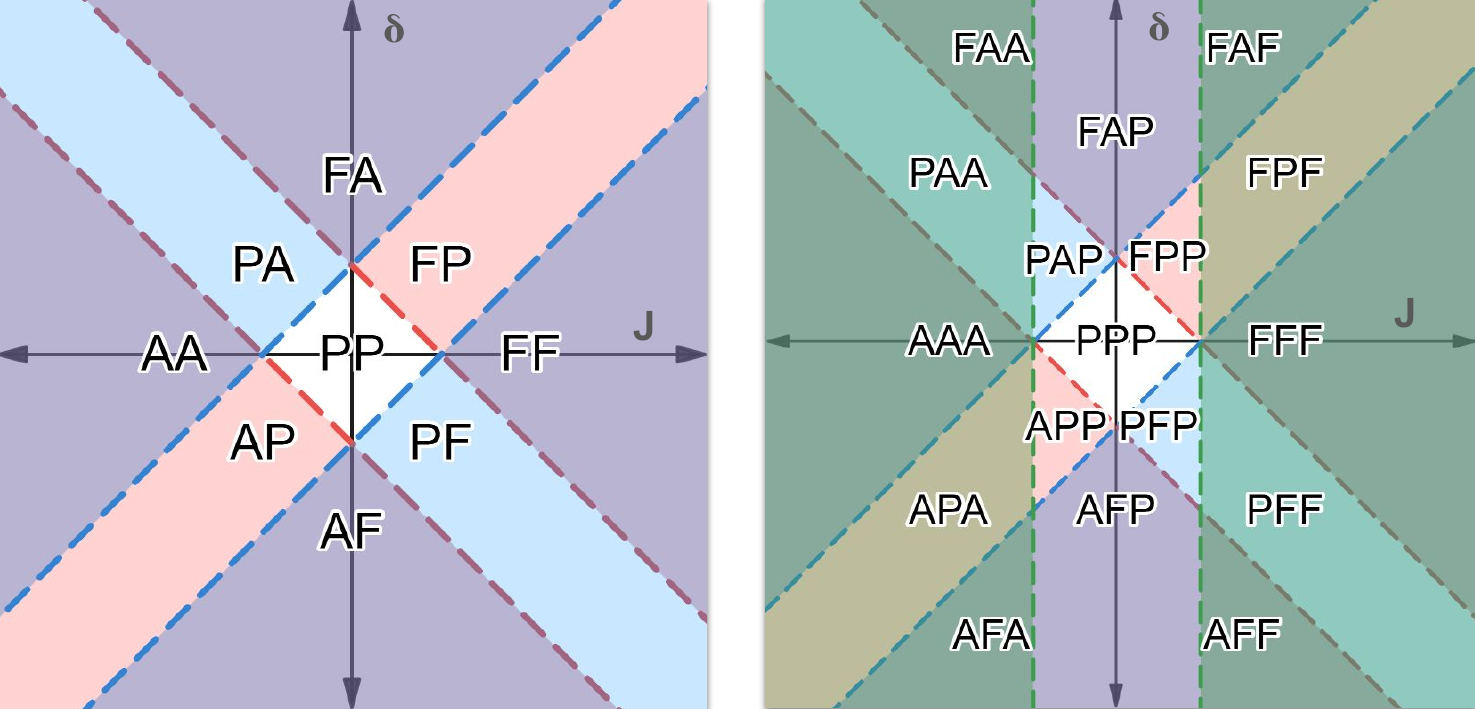}};

        \node[anchor=north west, font=\bfseries]
            at ([xshift=2mm,yshift=-2mm]img.north west) {(b)};

        \node[anchor=north west, font=\bfseries]
            at ([xshift=2mm,yshift=-2mm]img.north) {(c)};
    \end{tikzpicture}
\end{minipage}

\caption{Schematic of the setup for preparing superpositions of quantum Ising ground states and the corresponding superposition sectors.
(a) A one-dimensional transverse-field Ising chain coupled to a central spin through the quantum interaction defined in Eqs.~\eqref{eq:qint1}~and~\eqref{eq:qint2}.
(b) Superposition sectors in the $J$-$\delta$ plane for a spin-$1/2$ central system. The first and second letters denote the phases of the ground states $|g(J-\delta)\rangle$ and $|g(J+\delta)\rangle$, respectively.
(c) Superposition sectors in the $J$-$\delta$ plane for a spin-$1$ central system. The first, second, and third letters denote the phases of the ground states $|g(J+\delta)\rangle$, $|g(J-\delta)\rangle$, and $|g(J)\rangle$, respectively.
In panels (b) and (c), P, F, and A denote the paramagnetic, ferromagnetic, and anti-ferromagnetic phases, respectively. For panels (b) and (c), both axes are in units of transverse-field strength $h$. }
\label{fig:schematic}
\end{figure*}

The remainder of the paper is organized as follows. We present the many-body central spin model and the state-preparation protocol in Sec.~\ref{Sec2}. In Sec.~\ref{sec:3}, we investigate the magnetization and nearest-neighbor correlations across the different superposition sectors. We also examine the entanglement structure of the generated states and their sensitivity to the Ising interaction strength. Finally, we present our concluding remarks in Sec.~\ref{Sec4}.

{
\section{Model and State-preparation Protocol}
\label{Sec2}
We consider a one-dimensional transverse-field Ising chain coupled to a central auxiliary spin, which is taken to be either a spin-$1/2$ or a spin-$1$ system. We promote the usual interaction to a \emph{quantum interaction} $\mathbf{\hat{J}}$. 
The Hamiltonian of the total system is given by
\begin{equation}
\hat H=
-\sum_{j=1}^{L}\left[\mathbf{\hat{J}}
\sigma_j^x\sigma_{j+1}^x
+
h\sigma_j^z\right].
\label{eq:qint1}
\end{equation}
Here, the chain consists of $L$ spin-$1/2$ parties, arranged in a ring, with periodic boundary conditions, i.e., $\sigma_{L+1}^\alpha = \sigma^\alpha_{1}$.  
As shown schematically in Fig.~\ref{fig:schematic}(a), the quantum interaction $\mathbf{\hat{J}}$ 
is an operator that acts on the Hilbert space of the central spin. It is defined as,
\begin{equation}
\mathbf{\hat{J}}=J\mathbb{I}_C+\Delta(t) \hat{\mathcal{Z}}_C, 
\label{eq:qint2}
\end{equation}
where $J$ denotes the nearest-neighbor interaction coupling among the system on the ring, whereas $\Delta(t)$ is the coupling of the particles with the central spin $C$. For a spin-$1/2$ central system, we take
\begin{align}
    \hat{\mathcal{Z}}_C = \sigma^z_C,
\end{align}
whereas for a spin-$1$ central system,
\begin{align}
    \hat{\mathcal{Z}}_C = \Sigma_C^z = |+1\rangle\langle +1| - |-1\rangle\langle -1|.
\end{align}
Let $|m_C\rangle$ denote the eigenvector of $\hat{\mathcal{Z}}_C$, such that $\hat{\mathcal{Z}}_C|m_C\rangle = m|m_C\rangle$, where $m = -1, +1$ for spin-$1/2$ central system and $m = -1, 0 , +1$ for spin-$1$ central system. 

We observe that $[\hat{H}, \hat{\mathcal{Z}}_C] = 0$. This implies  that the total Hamiltonian $\hat{H}$ can be written in the block-diagonal form
\begin{align}
    \hat{H}(t) = \sum_m |m_C\rangle\langle m_C| \otimes \hat{H}_m(t),
    \label{Hamil:tot}
\end{align}
where $\hat{H}_m(t)$ is the Hamiltonian corresponding to the $m$th sector, acting on the chain with $L$-spins in a ring. It is given by
\begin{align}
    \hat{H}_m(t) =  -\sum_{j=1}^L \left[J_m(t)\sigma^x_j\sigma^x_{j+1} + h \sigma_j^z\right],
\end{align}
with $J_m(t) = J + m\Delta(t)$.

In order to prepare a superposition of quantum Ising ground states in different phases, we need to let the total system evolve while changing the parameter $\Delta(t)$ adiabatically. Initially, the quantum interaction coupling is switched off, i.e., $\Delta(0) = 0$. Initially, the quantum Ising chain is prepared in its ground state while the central spin is prepared in a superposition state $|\chi_C\rangle$. The coupling of the quantum interaction $\Delta(t)$ is adiabatically increased and set to a value $\Delta(T) = \delta$. {From this point onward, we rescale the interaction strengths and express the parameters $J$ and $\delta$ in units of the local transverse-field strength $h$.}

Initially, we have $\hat H_m(0)$
and the chain is prepared in the corresponding ground state $|g(J)\rangle$, while the central spin is prepared in $|\chi_C \rangle = \sum_m \tilde c_m |m_C\rangle$. Thus, the joint initial state is given by
\begin{align}
    |\Psi(0)\rangle = \sum_m \tilde c_m|m_C\rangle\otimes|g(J)\rangle.
\end{align}
Now, the state is evolved with the Hamiltonian $\hat H(t)$ as given in Eq.~\eqref{Hamil:tot}. The parameter $\Delta(t)$ is adiabatically changed from $\Delta(0) = 0$ to $\Delta(t) = \delta$, where $T$ is total time of the evolution. Consequently, the state evolves, and at time $t$, the state is given by
\begin{align}
    |\Psi(t)\rangle = \sum_m \tilde c_m |m_C\rangle \otimes \hat U_m(t) | g(J) \rangle,
\end{align}
where the $m$th sector evolution operator is given by
\begin{align}
    \hat U_m(t) = \mathcal{T} \exp\left[-\frac{i}{\hbar} \int_0^t ds \hat H_m(s)\right]. 
\end{align}
Here, $\mathcal{T}$ denotes time-ordering. Since the evolution is adiabatic, the eigenstates of the initial Hamiltonian follow the change in the Hamiltonian and evolve to become the corresponding eigenstates of the instantaneous Hamiltonian. The initial ground state evolves to become the ground state of the Hamiltonian $\hat H_m(t)$, i.e., $\hat U_m(t)|g(J)\rangle = e^{i\phi_m}|g(J+m\Delta(t))\rangle$. Here, $\phi_m$ is the phase accumulated by the $m$th sector adiabatic evolution. Thus, the evolved state at time $t = T$ is given by 
\begin{align}
    |\Psi(T)\rangle = \sum_m c_m|m_C\rangle\otimes |g(J + m\delta)\rangle.
\end{align}
Here, we have set $c_m = \tilde c_me^{i\phi_m}$. Following the state preparation, we measure the central spin in a basis mutually unbiased to the computational basis, and post-select the outcome $|\eta\rangle = \frac{1}{\sqrt d} \sum_m |m_C\rangle$, where $d = 2$ for spin-$1/2$ and $d = 3$ for spin-$1$ central system.
Consequently, we obtain the state in the chain as
\begin{align}
    |\psi_d\rangle = \frac{1}{\sqrt {\mathcal{N}}} \sum_m c_m |g(J+m\delta)\rangle.
\end{align}
The probability of obtaining $|\eta\rangle$, $$P_\eta = \frac{\mathcal{N}}{d} = \frac{1}{d}\sum_{m,n} c_m^* c_n \langle g(J+m\delta)| g(J+n\delta)\rangle.$$

In the thermodynamic limit, the transverse-field Ising model is in the paramagnetic phase (P) for \(|J_m|<h\). For \(|J_m|>h\), the system enters an ordered phase: the ground state is ferromagnetic (F) for \(J_m>0\) and anti-ferromagnetic (A) for \(J_m<0\). The quantum critical points separating these phases occur at \(J_m=\pm h\).

Now, if we consider the central system to be spin-$1/2$, then it is possible to prepare superpositions of two different phases for different values of $J$ and $\delta$. 
The explicit post-measurement state obtained is
\begin{equation}
|\psi_2\rangle=
    \frac{1}{\sqrt{\mathcal N}}
\left[{c_{+1} |g(J+\delta)\rangle+c_{-1}|g(J-\delta)\rangle}\right]
,
\label{eq:fin-qubit-state}
\end{equation}
with
$\mathcal N=|c_{+1}|^2+|c_{-1}|^2+2\text{Re}[c_{+1}^* c_{-1}F^{+-}(J,\delta)]$,
where $F^{+-}(J,\delta)=\langle g(J+\delta)|g(J-\delta)\rangle$. The various possible superposition sectors are presented in Fig.~\ref{fig:schematic}(b), where $h$ is set to be of unit value. We observe that when $J$ and $\delta$ are both small, such that $|J\pm \delta|<h$, both the states in superposition are the ground states corresponding to the paramagnetic phase, represented by PP. We observe that there are superposition sectors like AF, sectors where one of the states is in anti-ferromagnetic phase and the other in the ferromagnetic phase. These superpositions were not possible in the \emph{quantum field} protocol considered in Ref.~\cite{Rams2012}, but are possible due to the quantum interaction $\hat{\mathbf{J}}$ considered in this work. We further investigate various observables, entanglement measures and quantum Fisher information corresponding to such states in this work.

Now, for an arbitrary observable $\hat O$, the expectation value of $\hat O$ on the state $|\psi_2\rangle$ is
\begin{equation}
\langle\hat O\rangle= {\mathcal N}^{-1}\left(
|c_{+1}|^2O_{++}
+
|c_{-1}|^2O_{--}
+
2\text{Re}
\left[c_{+1}^* c_{-1}
O_{+-}
\right]\right),
\label{eq:avg-O2}
\end{equation}

where $O_{ab}=\langle g(J+a\delta)|\hat O|g(J+b\delta)\rangle$ with $a,b\in\{+1,-1\}$.}


{Similarly, for the central system with a spin-$1$, the state obtained after the measurement  is




\begin{equation}
|\psi_3\rangle=
\frac{1}{\sqrt{\mathcal N}}\left[
c_{+1}|g(J+\delta)\rangle
+
c_0|g(J)\rangle
+
c_{-1}|g(J-\delta)\rangle\right],
\label{eq:state}
\end{equation}
where
$\mathcal N= \sum_{a,b} c_a^*c_b F^{ab}$.
The overlaps are $F^{ab} = \langle g(J+a\delta)|g(J+b\delta)\rangle$, with $a,b \in \{+1, 0,-1\}$. For brevity, \(+1\) and \(-1\) are denoted by \(+\) and \(-\), respectively, in the overlap superscripts.
The expectation value of an arbitrary observable $\hat O$ is
\begin{equation}
\begin{split}
\langle \hat O\rangle
=&
{\mathcal N}^{-1}
\Big[
|c_{+1}|^2 O_{++}
+ |c_0|^2 O_{00}
+ |c_{-1}|^2 O_{--}  \\
&+2\mathrm{Re}
\Big(
c_{+1}^*c_0 O_{+0}
+ c_{+1}^*c_{-1} O_{+-}
+ c_{-1}^*c_{0} O_{-0}
\Big)
\Big].
\end{split}
\label{eq:avg-O3}
\end{equation}
where
$O_{ab}=
\langle g(J+a\delta)|\hat O|g(J+b\delta)\rangle$, with
$a,b\in\{+1,0,-1\}$. Again, we denote $+1$ and $-1$ as $+$ and $-$ in $O_{ab}$ for brevity.

The prepared state is a coherent superposition of three quantum Ising ground states corresponding to different interaction strengths. Depending on these strengths, the constituent ground states can belong to various combinations of the anti-ferromagnetic~(A), ferromagnetic~(F), and paramagnetic~(P) phases, as illustrated in Fig.~\ref{fig:schematic}(c). In particular, our spin-$1$-controlled interaction enables a coherent superposition involving all three phases, extending the two-component superpositions considered in the quantum-field protocol of Ref.~\cite{Rams2012}. We investigate the magnetization, spin correlations, and entanglement associated with these superposition sectors in the following sections.}
\begin{figure}[t]
\centering
\includegraphics[width=0.65\linewidth]{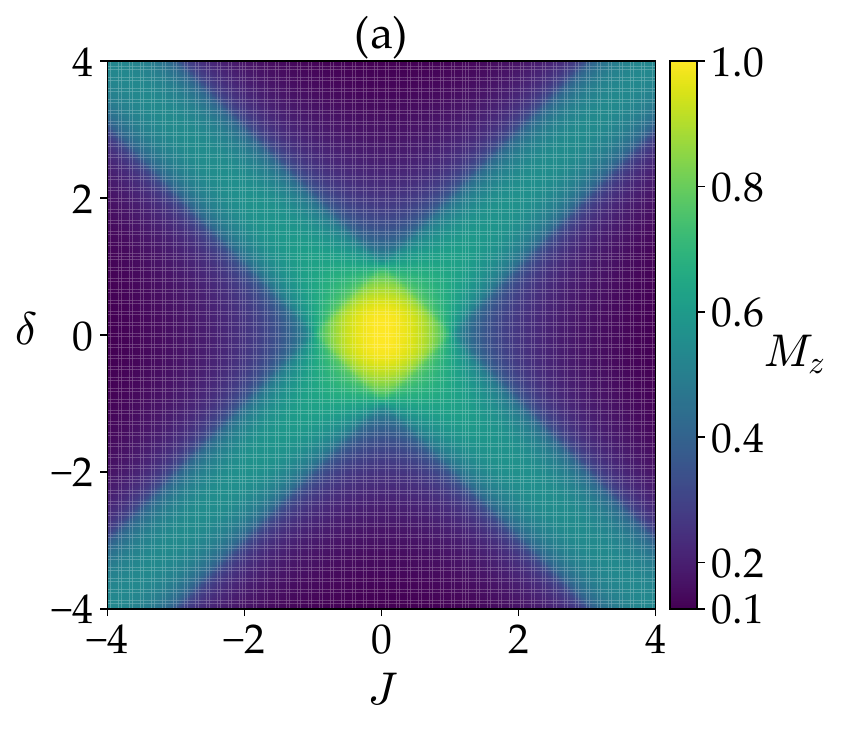}
\includegraphics[width=0.65\linewidth]{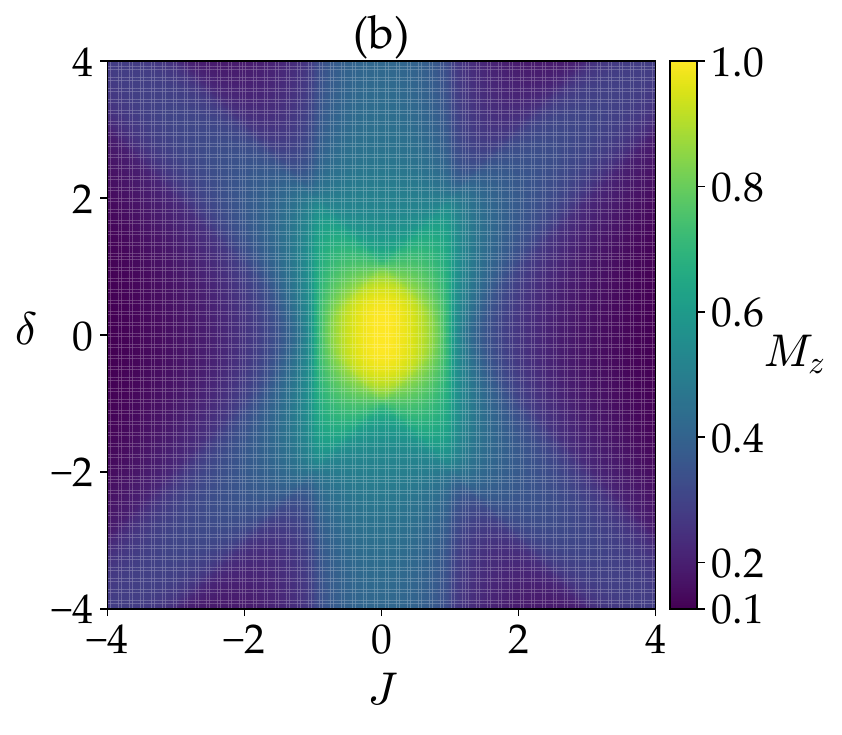}
\caption{Transverse magnetization $M_z$ of superpositions of quantum Ising ground states across the $J$--$\delta$ plane.
(a) The state given in Eq.~\eqref{eq:fin-qubit-state}, generated using a spin-$1/2$ central system, with $c_{+1}=c_{-1}=1/\sqrt{2}$.
(b) The state given in Eq.~\eqref{eq:state}, generated using a spin-$1$ central system, with $c_{+1}=c_0=c_{-1}=1/\sqrt{3}$. We have set $L=100$ here. For both panels, $x$ and $y$ axes are in units of transverse-field strength $h$, while the magnetization is unitless.} 
\label{fig:Z_qubit_heatmap}
\end{figure}

{
\section{Properties of the Superposition States}
\label{sec:3}
In this section, we investigate the magnetization and spin correlations of the post-selected states across the different superposition sectors. We then characterize their entanglement and examine their sensitivity to the interaction strength through the quantum Fisher information.
\\

\subsection{Magnetization and Correlations}
\label{sec:mag-corr}

\begin{figure*}[t]
\centering
\includegraphics[width=0.32\textwidth]{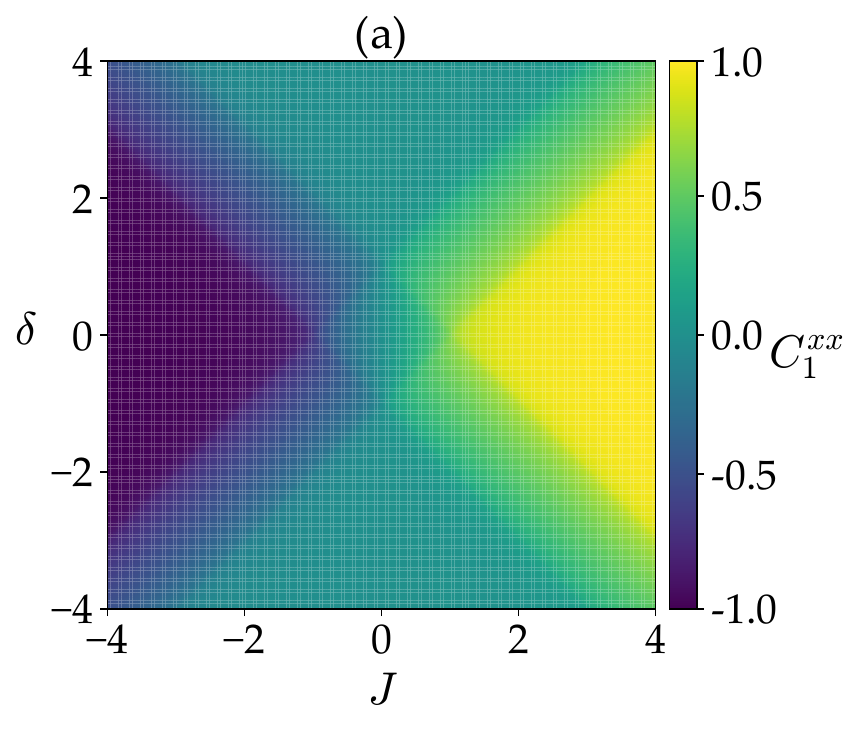 }
\includegraphics[width=0.32\textwidth]{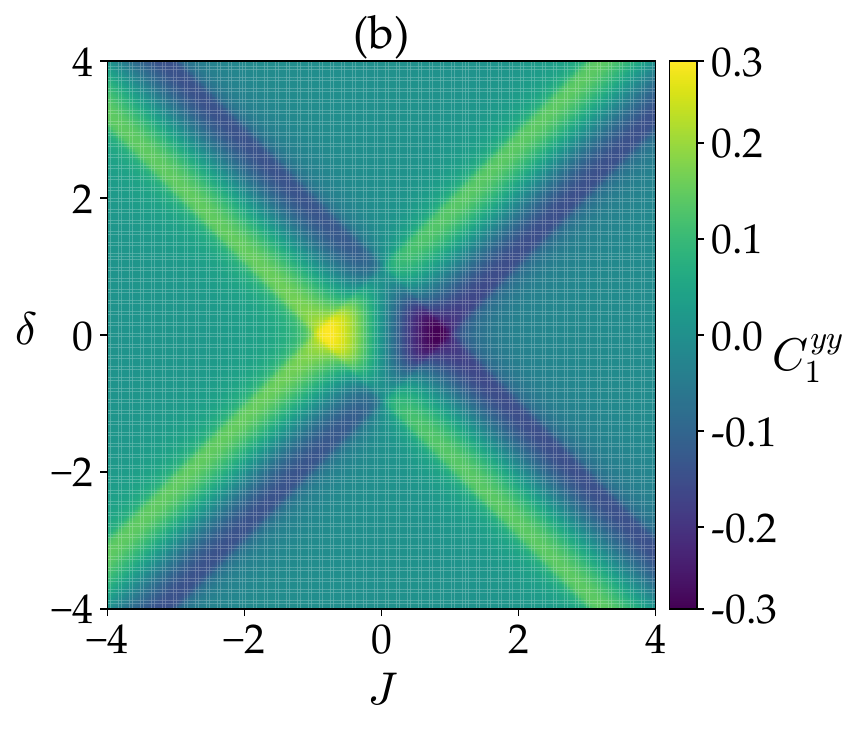 }
\includegraphics[width=0.32\textwidth]{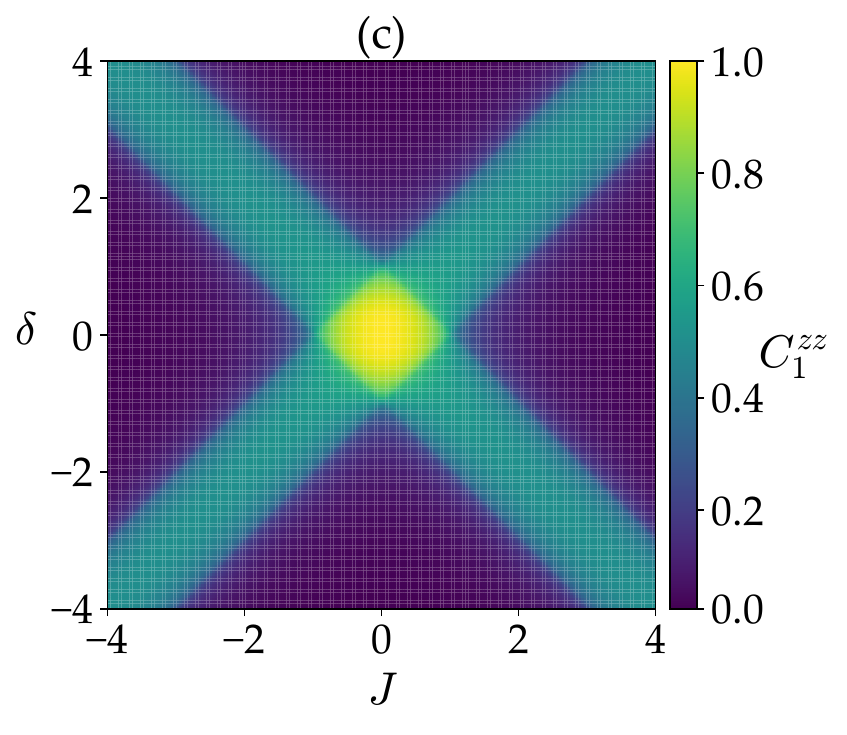 }
\caption{Nearest-neighbor spin correlations of the state given in Eq.~\eqref{eq:fin-qubit-state}, generated using a spin-$1/2$ central system, across the $J$--$\delta$ plane. Panels (a), (b), and (c) show $C_1^{xx}$, $C_1^{yy}$, and $C_1^{zz}$, respectively. The system size is $L=100$, with $c_{+1}=c_{-1}=1/\sqrt{2}$. For all the panels, $x$ and $y$ axes are in units of transverse-field strength $h$, while the correlations are unitless.}
\label{fig:XX_qubit_heatmap}
\end{figure*}

We first investigate the magnetization and nearest-neighbor correlations of the superposition states obtained in  Eqs.~\eqref{eq:fin-qubit-state} and \eqref{eq:state}. The magnetization along direction $\alpha$ is defined as
\begin{align}
M_\alpha = \frac{1}{L}\sum_{j=1}^{L}\langle \psi_I|\sigma^\alpha_j|\psi_I\rangle,
\label{eq:mag}
\end{align}
where $\alpha \in \{x,y,z\}$, $L$ is the total number of spins in the chain, and $I \in \{2,3\}$ denotes the states obtained by having the central system of spin-$1/2$ and spin-$1$, respectively. Similarly, the spin correlations at separation $r$ are defined as
\begin{align}
C^{\alpha\beta}_r = \frac{1}{L}\sum_{j=1}^{L}\langle \psi_I|\sigma^\alpha_j\otimes \sigma_{j+r}^\beta|\psi_I\rangle,
\label{eq:corr}
\end{align}
where $\alpha, \beta \in \{x,y,z\}$, $1 \leq r\leq L-1$. For the translationally invariant states considered here, the expectation values in Eqs.~\eqref{eq:mag}~and~\eqref{eq:corr} are independent of $j$. We can therefore evaluate each quantity at a single site, which we choose as $j=1$. Throughout this work, we restrict our analysis to nearest-neighbor correlations, corresponding to $r=1$. As follows from Eqs.~\eqref{eq:avg-O2}~and~\eqref{eq:avg-O3}, evaluating these observables requires both the diagonal matrix elements and the cross terms between the ground states forming each superposition. 
}

{Following the methods of Refs.~\cite{Pfeuty1970, Barouch1971, Mbeng2024}, we evaluate the magnetization $M_z$ using the matrix
\begin{equation}
\langle \sigma_j^z\rangle_{ab}=F^{ab}\left[1-2\sum_{k>0}\mathcal{X}_k^{ab}\right],
\label{eq:mag-cross}
\end{equation}
where $\langle\sigma_j^z\rangle_{ab} = \langle g(J+a\delta)|\sigma_j^z|g(J+b\delta)\rangle$ $\mathcal{X}^{ab}_k$ is defined in Eq.~\eqref{eq:X-mag}. 
The magnetizations along the $x$ and $y$ directions vanish, yielding $M_x=M_y=0$, as shown in Appendix~\ref{zero}.

Similarly, the only non-zero nearest-neighbor spin correlations are given by
\begin{align}
\langle \sigma_j^x\sigma_{j+1}^x\rangle_{ab}&=F^{ab}\sum_{k>0}\Big[2\mathcal{X}_k^{ab}\cos k +\left(\mathcal{Y}_k^{ab}+\mathcal{Y}_k^{ba}\right)\sin k\Big], \\
\langle\sigma_j^y\sigma_{j+1}^y
\rangle_{ab}
&=F^{ab}\sum_{k>0}\Big[2\mathcal{X}_k^{ab}\cos k  -\left(\mathcal{Y}_k^{ab} + \mathcal{Y}_k^{ba}\right)\sin k\Big],
\end{align}
\begin{widetext}
\begin{equation}
\langle
\sigma_j^z
\sigma_{j+1}^z
\rangle_{ab}
=
4F^{ab}
\Bigg[
\frac{1}{4}
-
\sum_{k>0} \mathcal{X}_k^{ab}
+
\left(
\sum_{k>0}
\mathcal{X}_k^{ab}
\right)^2
-
\left(\sum_{k>0} \mathcal{X}_k^{ab} \cos k \right)^2
+
\left(\sum_{k>0} \mathcal{Y}_k^{ab} \sin k\right)
\left(
\sum_{l>0}
\mathcal{Y}_l^{ba} \sin l\right)
\Bigg].
\end{equation}
\end{widetext}
Here, $\langle\sigma^\alpha_j\sigma_{j+1}^\alpha\rangle_{ab} = \langle g(J+a\delta)|
\sigma_j^\alpha
\sigma_{j+1}^\alpha
|g(J+b\delta)\rangle$, with $\alpha \in \{x,y,z\}$. 
The explicit forms of $\mathcal{X}_k^{ab}$, $\mathcal{Y}_k^{ab}$, and $\mathcal{Y}_k^{ba}$ are presented in Eq.~\eqref{eq:X-mag}, and $F^{ab}$ is given in Eq.~\eqref{eq:cross-fidelity}. As discussed earlier $a,b \in \{-1, 0, +1\}$ when the central system is spin-$1$, while for the case of spin-$1/2$, $a,b \in \{-1, +1\}$. A detailed derivation is presented in Appendix~\ref{AppB}.
}



\begin{figure*}[t]
\centering
\includegraphics[width=.33\textwidth]{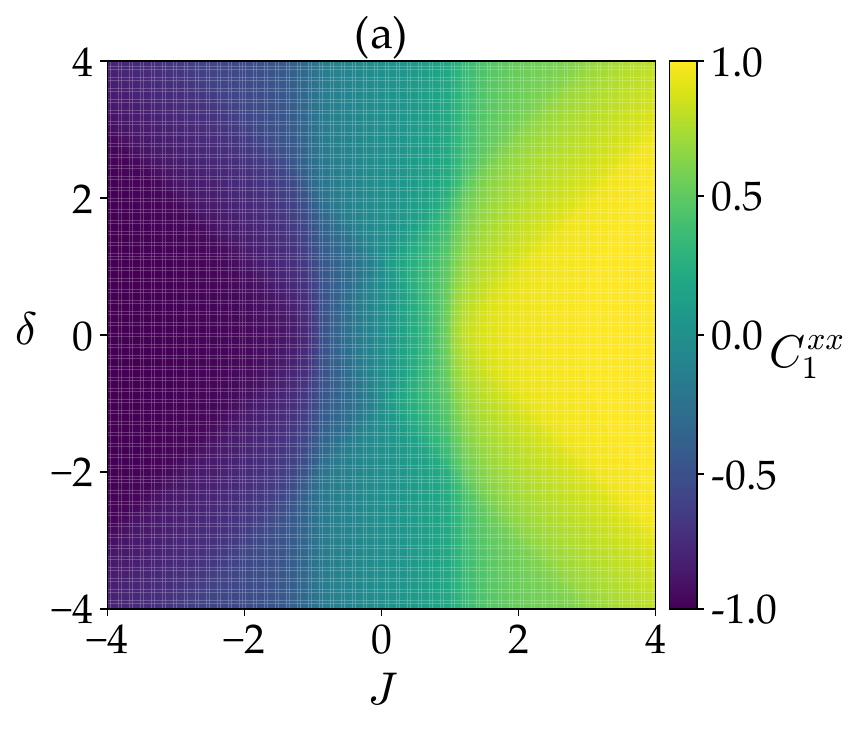 }
\includegraphics[width=.33\textwidth]{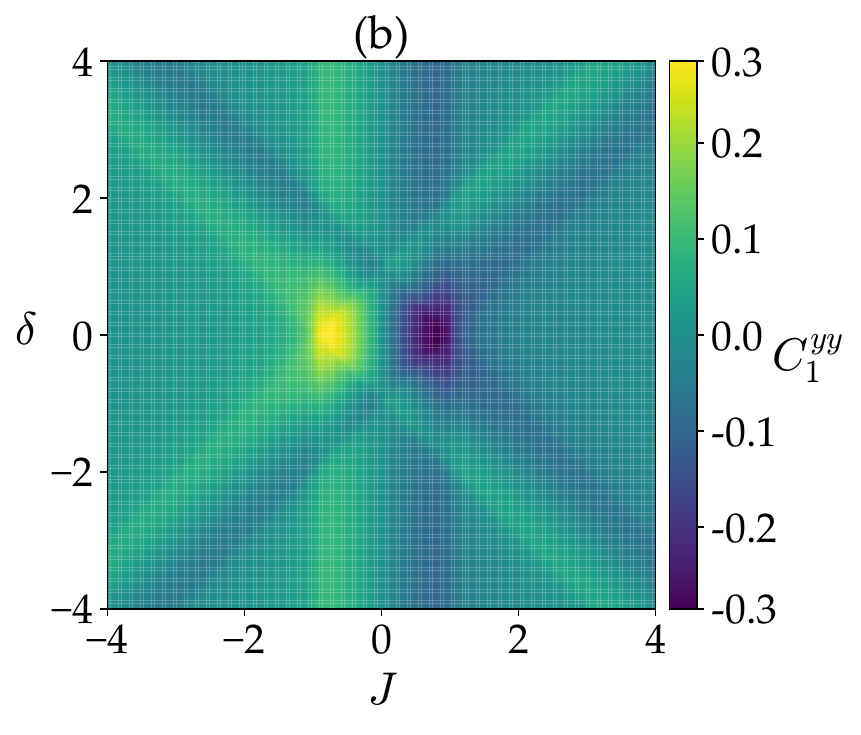 }
\includegraphics[width=.326\textwidth]{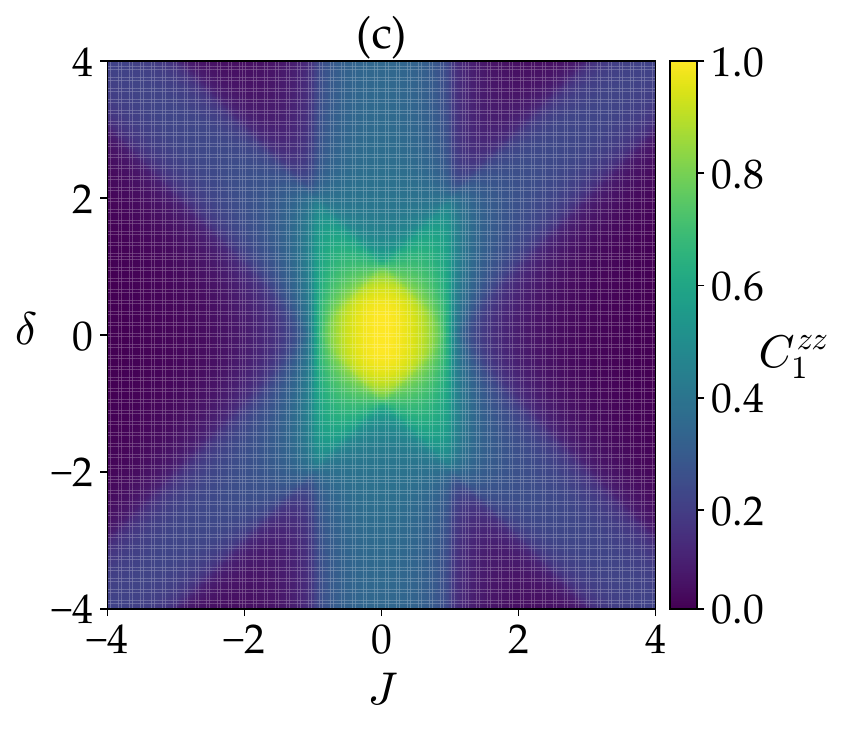 }
\caption{Nearest-neighbor spin correlations of the state given in Eq.~\eqref{eq:state}, generated using a spin-$1$ central system, across the $J$--$\delta$ plane. Panels (a), (b), and (c) show $C_1^{xx}$, $C_1^{yy}$, and $C_1^{zz}$, respectively. The system size is $L=100$, with $c_{+1}=c_0=c_{-1}=1/\sqrt{3}$. For all the panels, $x$ and $y$ axes are in units of transverse-field strength $h$, while the correlations are unitless.}
\label{fig:XX_qutrit_heatmap}
\end{figure*}

{
Figure~\ref{fig:Z_qubit_heatmap} shows the transverse magnetization $M_z$ as a function of $J$ and $\delta$ for the states $|\psi_2\rangle$ and $|\psi_3\rangle$, defined in Eqs.~\eqref{eq:fin-qubit-state} and \eqref{eq:state}, respectively. Here, all the ground states are in equal superposition. The magnetization patterns closely reflect the corresponding superposition sectors shown in Figs.~\ref{fig:schematic}(b) and (c). In the transverse-field Ising chain, a dominant transverse field favors spin alignment along $z$, resulting in a large $M_z$, whereas stronger ferromagnetic or anti-ferromagnetic interactions reduce the transverse magnetization. This trend is also evident in the superposition states considered here. The PP and PPP sectors exhibit the highest magnetization for the spin-$1/2$ and spin-$1$ central systems, respectively. Sectors containing both paramagnetic and ordered ground-state components exhibit intermediate values, while those composed entirely of ferromagnetic and/or anti-ferromagnetic components have the lowest magnetization.
The differences between the results with the central system being a spin-$1/2$ and a spin-$1$ can be understood from the interaction strengths associated with their constituent ground states. In the spin-$1/2$ case, the components change phase along the lines $J\pm\delta=\pm h$. The spin-$1$ superposition additionally contains the component $|g(J)\rangle$, whose phase boundaries occur at $J=\pm h$, independently of $\delta$. Consequently, this component remains paramagnetic throughout the strip $|J|<h$, even when the two shifted components belong to the ferromagnetic and anti-ferromagnetic phases. Its contribution explains the persistence of appreciable transverse magnetization $M_z$ near $J=0$ in the spin-$1$ case, where the spin-$1/2$ magnetization is suppressed for large $|\delta|$. The additional vertical structure in the spin-$1$ heatmaps thus reflects the presence of the unshifted ground-state component.
}


{ Now, we consider the nearest-neighbor correlations of the equal superposition states of the form of Eqs.~\eqref{eq:fin-qubit-state} and \eqref{eq:state}. We first note that there are some symmetries in the magnetization and correlation plots. For the equal-weight real superpositions considered here, changing $\delta$ to $-\delta$ only keeps the states under consideration invariant and therefore leaves the observables unchanged. 
We observe that the quantities $C_1^{xx}$ and $C_1^{yy}$ are odd functions in $J$, so they change their sign under $J\to -J$ and are zero at $J = 0$. On the other hand, $M_z$ and $C_1^{zz}$ remain unchanged under $J\rightarrow -J$.
This occurs even when the shifted components belong to ferromagnetic and anti-ferromagnetic phases, as their contributions cancel.}

{In Figs.~\ref{fig:XX_qubit_heatmap}(a) and \ref{fig:XX_qutrit_heatmap}(a), we present the nearest-neighbor $XX$ correlation, $C_1^{xx}$. We observe large positive correlations in sectors dominated by ferromagnetic ground states. On the other hand, we find large negative correlations in those dominated by anti-ferromagnetic ground states. 
In the superposition sectors, where both types of ground states are present, their contributions have opposite signs and partially cancel. 
This leads to smaller magnitudes of $C_1^{xx}$ and a smooth change between positive and negative values. For the spin-$1$ case, the additional state $|g(J)\rangle$ also contributes, changing this behavior according to the phase to which it belongs. 

The $YY$ correlations, $C^{yy}_1$ are shown in Figs.~\ref{fig:XX_qubit_heatmap}(b) and \ref{fig:XX_qutrit_heatmap}(b). Compared to the $XX$ correlations, their magnitudes are smaller. They are more pronounced when the transverse field and the Ising interaction have comparable strengths, and become small when either of them dominates. 

The $ZZ$ correlations, shown in Figs.~\ref{fig:XX_qubit_heatmap}(c) and \ref{fig:XX_qutrit_heatmap}(c), follow a pattern similar to that of $M_z$. Their largest values occur in the purely paramagnetic sectors. The correlations decrease as the ferromagnetic or anti-ferromagnetic components become more prominent. In the spin-$1$ case, $|g(J)\rangle$ remains paramagnetic for $|J|<h$, irrespective of $\delta$. Thus, appreciable $ZZ$ correlations persist in this region even when the other two components belong to ordered phases. This explains the vertical feature around $J=0$ in the spin-$1$ plot.}

{
\subsection{Entanglement structure }

We now investigate the bipartite and multipartite entanglement of the superposition states $|\psi_2\rangle$ and $|\psi_3\rangle$, defined in Eqs.~\eqref{eq:fin-qubit-state} and \eqref{eq:state}, respectively. To quantify bipartite entanglement between two nearest-neighbor spins, we use concurrence~\cite{Hill1997,Wootters1998}. Concurrence is an entanglement monotone that ranges from $0$ for separable states to $1$ for maximally entangled two-spin-$1/2$ states. It is monotonically related to the entanglement of formation, defined as the minimum average pure-state entanglement over all decompositions of the density matrix.

For a two-spin-$1/2$ density matrix $\rho$, the concurrence is given by
\begin{equation}
    \mathcal{C}=\max\{0,\sqrt{\lambda_1}-\sqrt{\lambda_2}-\sqrt{\lambda_3}-\sqrt{\lambda_4}\}
\end{equation}
where $\lambda$'s are eigenvalues of  $\rho\tilde{\rho}$ where $\tilde{\rho}=(\sigma_y\otimes\sigma_y)\rho^*(\sigma_y\otimes\sigma_y)$ and $\rho^*$ is complex conjugate of $\rho$, with $\lambda_1$ being the largest eigenvalue.
The reduced density matrix of spins $i$ and $j$ is obtained by tracing out the remaining spins and can be expanded in the Pauli basis as
\begin{equation}
\rho_{ij}
=\text{Tr}_{\overline{ij}}\left(|\psi_I\rangle\langle\psi_I|\right)
=\frac{1}{4}\sum_{\alpha,\beta=0}^{3}
p_{\alpha\beta} \, \sigma_i^\alpha\otimes\sigma_j^\beta,
\end{equation}
where $\text{Tr}_{\overline{ij}}$ denotes partial-trace over all sites except $i$ and $j$, and $\sigma^0=\mathbb{I}$, $\sigma^1=\sigma^x$, $\sigma^2=\sigma^y$, and $\sigma^3=\sigma^z$. The expansion coefficients are
\begin{equation}
p_{\alpha\beta}
=\text{Tr}\left[\rho_{ij}
\left(\sigma_i^\alpha\otimes\sigma_j^\beta\right)\right]
=\langle\sigma_i^\alpha\sigma_j^\beta\rangle.
\end{equation}
We restrict our analysis to nearest-neighbor concurrence, corresponding to $i=j$ and the second site being $j+1$ in the correlation notation. For the translationally invariant, real superpositions with fixed parity considered here, the nearest-neighbor reduced density matrix is completely determined by the magnetization $M_z$ and the correlations $C_1^{xx}$, $C_1^{yy}$, and $C_1^{zz}$. The remaining nontrivial spin expectation values vanish, as discussed in App.~\ref{zero}.}
\begin{figure}[h]
\centering
\includegraphics[width=0.65\linewidth]{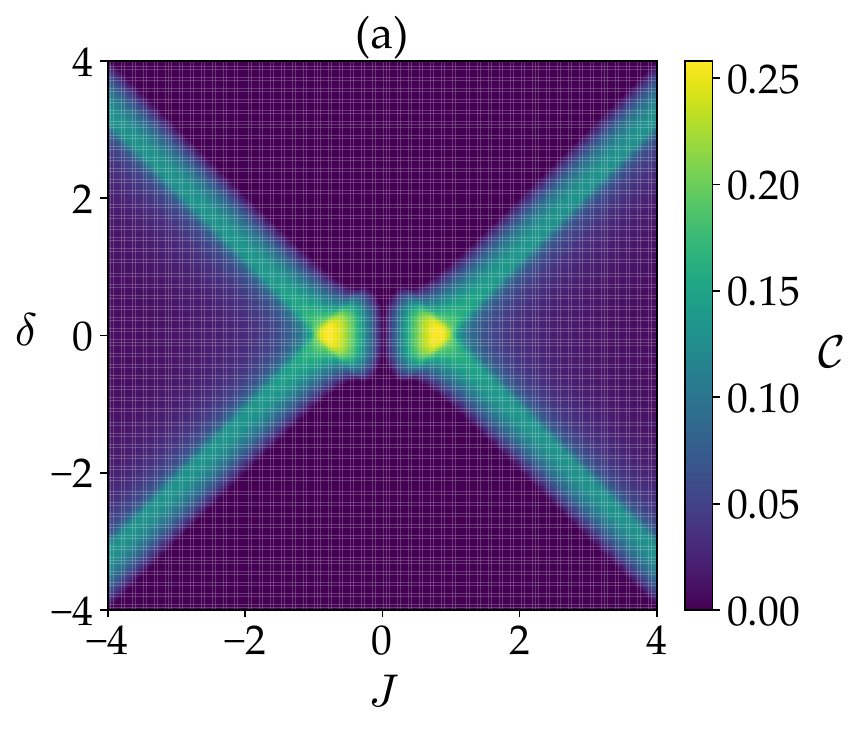}
\includegraphics[width=0.65\linewidth]{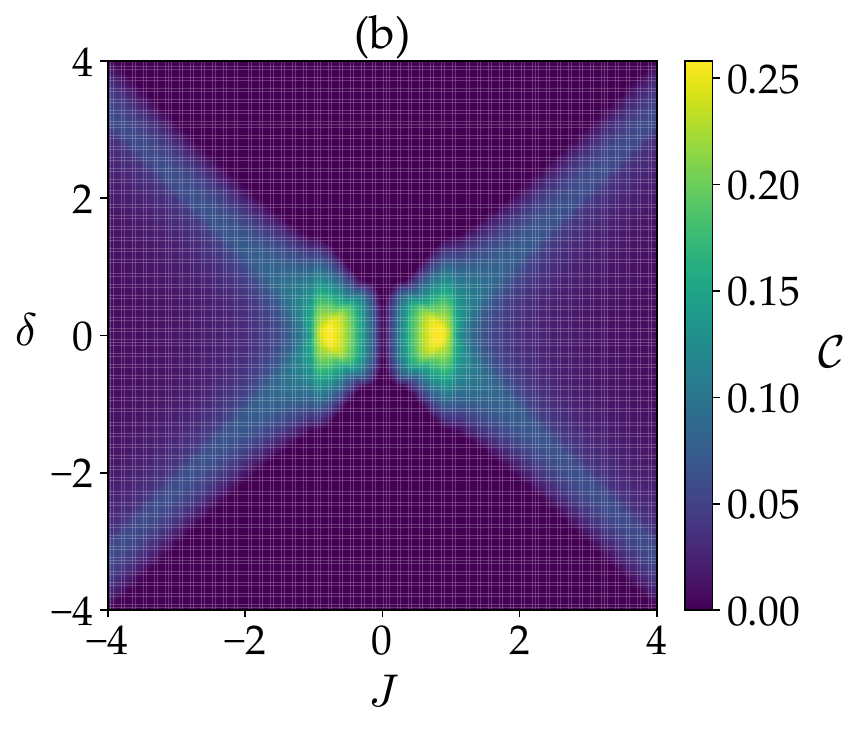 }
\caption{Nearest-neighbor concurrence $\mathcal{C}$ (in ebits) of superpositions of quantum Ising ground states with equal weights, across the $J-\delta$ plane.
(a) The state given in Eq.~\eqref{eq:fin-qubit-state}, generated using a spin-$1/2$ central system, with $c_{+1}=c_{-1}=1/\sqrt{2}$.
(b) The state given in Eq.~\eqref{eq:state}, generated using a spin-$1$ central system, with $c_{+1}=c_{-1}=c_0=1/\sqrt{3}$.
The system size is $L=100$ in both panels. For both panels, $x$ and $y$ axes are in units of transverse-field strength $h$.}
\label{fig:Concurrence_qubit_heatmap}
\end{figure}
{
The nearest-neighbor concurrence in the transverse-field Ising model has been studied in Refs.~\cite{Osborne2002,Osterloh2002}. Its maximum occurs on the paramagnetic side, at $|J_{\max}|<h$, close to but not exactly at the transition point. We observe a related pattern over the larger $(J,\delta)$ parameter space for the superposition states considered here. Figures~\ref{fig:Concurrence_qubit_heatmap}(a) and (b) show the concurrence for the spin-$1/2$ and spin-$1$ central systems, respectively, with equal amplitudes for all components. At $\delta=0$, the components coincide, and both superpositions reduce to the quantum Ising ground state $|g(J)\rangle$.

\begin{figure}[h]
\centering
\includegraphics[width=0.65\linewidth]{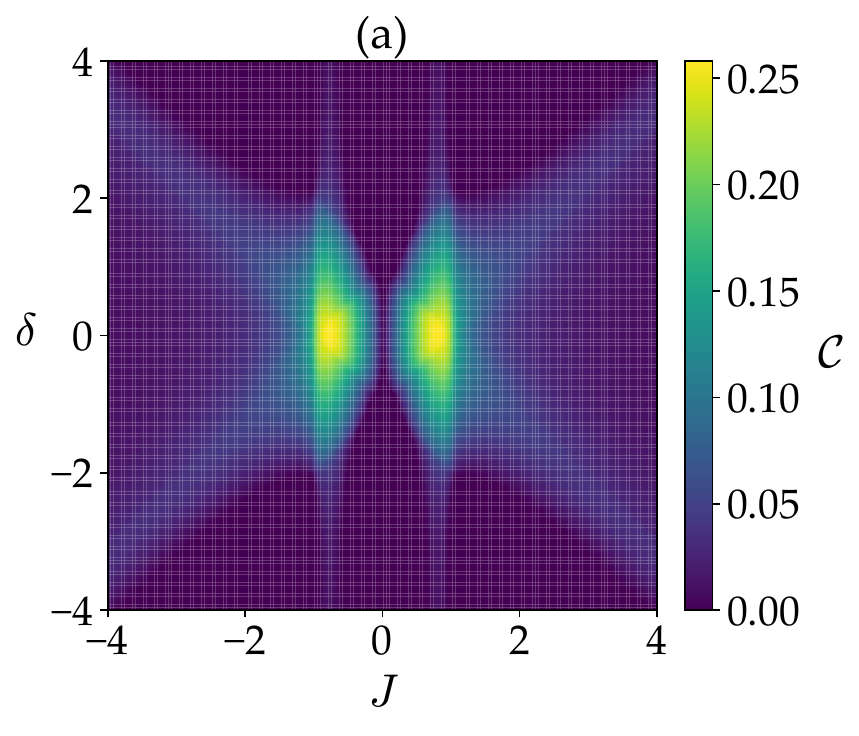 }
\includegraphics[width=0.65\linewidth]{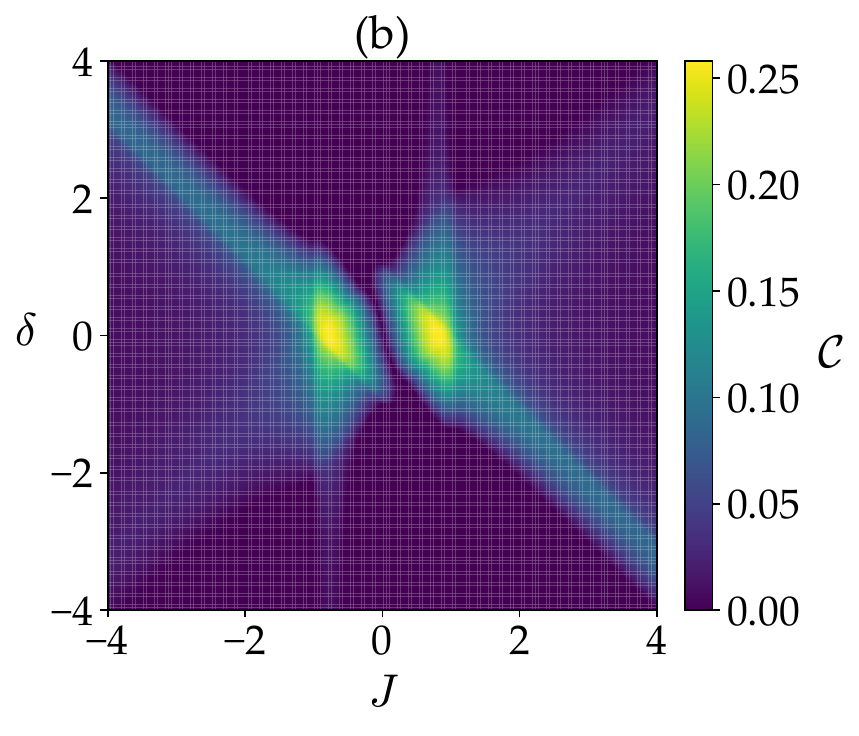}
\caption{Nearest-neighbor concurrence $\mathcal{C}$ (in ebits) of superpositions of quantum Ising ground states with unequal weights, across the $J$-$\delta$ plane. The state considered is given in Eq.~\eqref{eq:state}, generated using a spin-$1$ central system.
(a) The coefficients are $c_{+1}=c_{-1}=1/\sqrt{6}$,and $c_0=2/\sqrt{6}$.
(b) The coefficients are  $c_{+1}=c_0={2/3}$, and $c_{-1}={1/3}$ .
The system size is $L=100$ in both panels. For both panels, $x$ and $y$ axes are in units of transverse-field strength $h$.}
\label{fig:Concurrence_qutrit_heatmap_2,1,2}
\end{figure}

In Fig.~\ref{fig:Concurrence_qubit_heatmap}(a), the largest concurrence is concentrated near $\delta=0$, on the paramagnetic side of the critical points $J=\pm1$, where we have set $h=1$. Away from $\delta=0$, appreciable concurrence extends along diagonal bands through the mixed sectors containing a paramagnetic component, namely FP, PF, AP, and PA. The concurrence varies within these sectors, while it is generally small deep inside the sectors containing only ferromagnetic and anti-ferromagnetic components.
Figure~\ref{fig:Concurrence_qubit_heatmap}(b) shows a similar overall pattern for the spin-$1$ case. The largest values again occur near $\delta=0$, while the diagonal bands are less pronounced than in the spin-$1/2$ case. Thus, in both cases, the regions of enhanced nearest-neighbor entanglement lie mainly within parts of the sectors containing paramagnetic components, rather than being uniform throughout these sectors.
 } 


{
We now consider unequal amplitudes of the three ground-state components, as shown in Fig.~\ref{fig:Concurrence_qutrit_heatmap_2,1,2}. In panel (a), we choose $c_{+1}=c_{-1}=1/\sqrt{6}$ and $c_0=2/\sqrt{6}$. Thus, the state has a larger weight to the component $|g(J)\rangle$. Compared with the equal-amplitude case, the central regions of enhanced concurrence become more extended along $\delta$. On the other hand, the diagonal bands are weaker. Since the two shifted components have equal amplitudes, the plot remains symmetric under both $J\rightarrow-J$ and $\delta\rightarrow-\delta$.

In Fig.~\ref{fig:Concurrence_qutrit_heatmap_2,1,2}(b), we consider $c_{+1}=c_0={2/3}$ and $c_{-1}={1/3}$. The larger amplitude of $|g(J+\delta)\rangle$ makes the diagonal band passing through the PAA and PFF sectors more pronounced. In these sectors, this component is paramagnetic, while the other two components are not. The band associated with a paramagnetic $|g(J-\delta)\rangle$ component is weaker. Thus, the unequal amplitudes remove the symmetries in $J$ and $\delta$ and change the distribution of nearest-neighbor entanglement across the sectors.

Thus, the concurrence patterns remain qualitatively similar for the spin-$1/2$ and spin-$1$ cases. But when the relative amplitudes are not equal, there are changes in the strength and extent of the entangled regions. The concurrence also behaves differently from the magnetization and spin correlations. The largest values of entanglement occur where the transverse field and the Ising interaction compete. These regions are near but not exactly at the critical points. It becomes small deep in the field-dominated and interaction-dominated regimes. Hence, large spin correlations do not necessarily imply large nearest-neighbor entanglement.}


{Along with the entanglement between two neighboring spins, we investigate the genuine multipartite entanglement of the prepared states. A pure state is genuinely multipartite entangled if it cannot be written as a product state across any bipartition of the system. To quantify this entanglement, we use the generalized geometric measure (GGM)~\cite{SenDe2010, Das2016}. This measure compares the given state with pure states that are separable across at least one bipartition. It is defined as one minus the maximum squared overlap with such states. Thus, the GGM vanishes for pure states without genuine multipartite entanglement and is positive otherwise. For an $N$-party pure state $|\phi_N\rangle$, it can be expressed in terms of the Schmidt coefficients as
\begin{equation}
\mathcal{G}(|\phi_N\rangle)
=
1
-
\max_{I:L}
\left\{
\xi_{I:L}^{\,2}
\right\},
\label{eq:GGM}
\end{equation}
where the maximization is taken over all bipartitions
\[
I \cup L = \{A_1,A_2,\ldots,A_N\},
\qquad
I \cap L = \varnothing,
\]
and $\xi_{I:L}$ is the largest Schmidt coefficient of
$|\phi_N\rangle$ corresponding to the bipartition $I:L$.}

{For the post-selected states considered here, our numerical calculations for system sizes $L=4,6,$ and $8$ show that the maximum in Eq.~\eqref{eq:GGM} is attained for a $1:\text{Rest}$ bipartition throughout the parameter range studied~\cite{Roy2019}. Since the states are translationally invariant, all single-site bipartitions give the same result. We therefore consider the reduced density matrix of the first spin,
\begin{equation}
    \rho_i=\Tr_{\overline{i}}(\rho)=\frac{1}{2}{\sum_{\alpha=0}^3q_{\alpha}\sigma_j^{\alpha}}
\end{equation} 
where $q_{\alpha}=\Tr(\sigma^{\alpha}_i\rho)=\langle\sigma^{\alpha}_i\rangle$.

As discussed in Sec.~\ref{sec:mag-corr}, only the local magnetization along the $z$ axis is nonzero. The single-site reduced density matrix therefore takes the form
\begin{equation}
\rho_1
=
\frac{1}{2}
\left(
\mathbb{I}+\langle \sigma^z\rangle \sigma^z
\right).
\end{equation}
Consequently, the largest eigenvalue is
$\xi_{\max}
=
({1+\langle \sigma^z\rangle})/{2}$.
Thus, the GGM becomes
\begin{equation}
\mathcal{G}
=
\frac{1-\langle \sigma^z\rangle}{2}.
\end{equation}

We thus find that the GGM is directly related to the transverse magnetization for the states considered here. Regions with strong transverse polarization have smaller GGM, while those with weaker polarization have larger GGM. Thus, magnetization and GGM show opposite trends across the different superposition sectors.}

{
\subsection{Quantum Fisher Information}

\begin{figure}[t]
    \centering
    \includegraphics[width=0.85\linewidth]{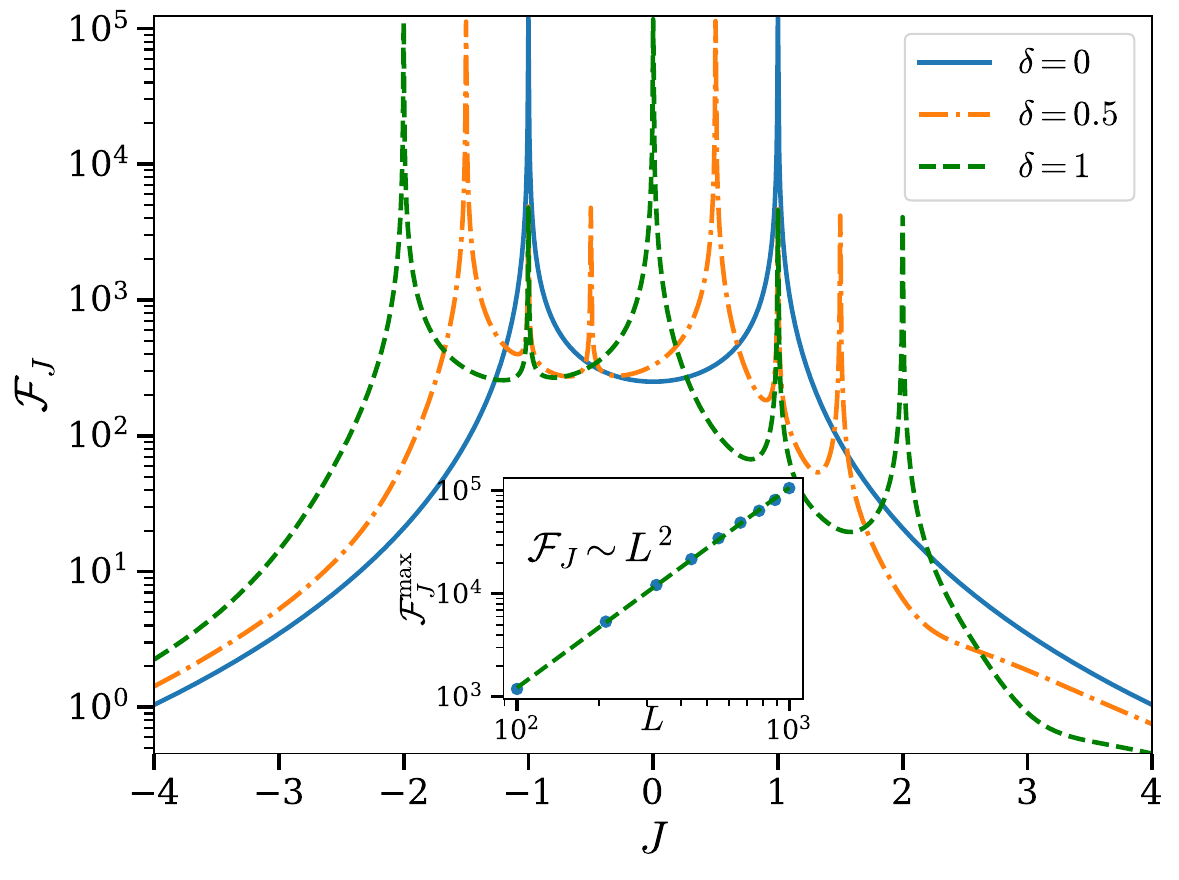}
    \caption{Quantum Fisher information $\mathcal{F}_J$ of the state given in Eq.~\eqref{eq:state} as a function of the Ising interaction strength $J$ (in units of h) for $\delta=0$ (solid-blue), $0.5$ (dot-dashed-orange), and $1$ (dashed-green). The system size is $L=100$, with $c_{+1}=5/\sqrt{27}$ and $c_{-1}=c_0=1/\sqrt{27}$. The inset shows the scaling of the maximum QFI with system size $L = 100$ to $1000$, for $\delta=1$, consistent with $\mathcal{F}_J^{\max}\sim L^2$. The $x$-axis is in units of transverse-field strength $h$, while the $y$-axis is in units of $h^{-2}$.}
    \label{fig:QFI_scaling}
\end{figure}

Quantum Fisher Information (QFI) measures how sensitively the quantum state changes if a parameter is changed. For $\nu$ independent repetitions, the variance of a locally unbiased estimator $\hat{\theta}$ of a parameter $\theta$ satisfies the quantum Cramér-Rao inequality~\cite{Braunstein94,PARIS2009,Sidhu2020,Helstrom74,Albarelli2020,Liu2019}
\begin{equation}
    \Delta^2\hat{\theta}
    \geq
    \frac{1}{\nu \mathcal{F}_Q(\theta)}.
\end{equation}
Here, $\mathcal{F}_Q(\theta)$ is the QFI. Thus, a larger QFI indicates greater sensitivity to the parameter and allows, in principle, more precise estimation.
For pure states, the QFI corresponding to the parameter $J$ is 
\begin{equation}
\mathcal{F}_J = 4\left[
\langle \partial_J\psi_I|\partial_J\psi_I\rangle
-
\left|\langle \psi_I|\partial_J\psi_I\rangle\right|^2
\right],
\end{equation}
where $|\psi_I\rangle$ is given by Eqs.~\eqref{eq:fin-qubit-state} and \eqref{eq:state}. The explicit expression for this state is given by
\begin{align}
    \mathcal{F}_J = 4\left[ \frac{\mathcal{A}}{\mathcal{N}} - \frac{|\mathcal{B}|^2}{\mathcal{N}^2}\right].
    \label{eq:qfi}
\end{align}
The explicit forms of $\mathcal{A}$ and $\mathcal{B}$ are given in Eqs.\eqref{eq:A} and \eqref{eq:B}. The derivation is presented in App.~\ref {App-QFI}.

For many-body spin systems, QFI has been studied extensively~\cite{ghosh2026, Agarwal_2025, Montenegro_2025}. Near a quantum critical point, the ground state becomes particularly sensitive to changes in the interaction parameters, leading to an enhanced QFI. To examine how this sensitivity depends on the system size, we consider the scaling of the maximum QFI,
\begin{equation}
    \mathcal{F}_J^{\max} \sim L^{\alpha}
\end{equation}
Taking the number of spins $L$ as the resource, linear growth, $\alpha=1$, corresponds to standard quantum-limit scaling, while quadratic growth, $\alpha=2$, corresponds to Heisenberg scaling.

In Fig.~\ref{fig:QFI_scaling}, we present the QFI as a function of $J$ for different values of $\delta$. For $\delta=0$, all three components coincide, and the normalized superposition reduces to a single quantum Ising ground state. We therefore observe two peaks near $J=\pm h$. For nonzero $\delta$, the components reach their critical points at different values of $J$, determined by $J+a\delta=\pm h$, where $a\in{-1,0,+1}$. This accounts for the additional peaks in the figure. For $h=1$, the peaks occur near $J=\pm0.5,\pm1,\pm1.5$ when $\delta=0.5$, and near $J=0,\pm1,\pm2$ when $\delta=1$. Their different heights reflect the unequal contributions of the ground-state components to the superposition.

The inset of Fig.~\ref{fig:QFI_scaling} shows the scaling of the maximum QFI with $L$ for $\delta=1$. We calculate the maximum QFI for system sizes $L = 100$ to $1000$. We observe that the maximum QFI scales with system size as $\mathcal{F}_J^{\max}\sim L^2$, indicating Heisenberg scaling near the critical regions. For a fixed number of successful post-selected preparations, the corresponding bound on the estimation uncertainty therefore decreases as $1/L$.}


\section{Conclusion}
\label{Sec4}
{In this work, we have investigated the superposition of ground states of the transverse-field Ising chain, prepared using a quantum interaction between the quantum Ising chain and a central-spin system. The states are produced by first adiabatically evolving the joint quantum Ising chain and central spin system, followed by projective measurement in the central-spin system and post-selection of the outcome. We are able to produce states that are a superposition of two or three ground states for a spin-$1/2$ or spin-$1$ central system, respectively. The resulting states can have components from ferromagnetic, anti-ferromagnetic, and paramagnetic phases.

We have studied the magnetization, nearest-neighbor correlations, and entanglement properties of these superposition states. The transverse magnetization and $ZZ$ correlations are largest in the purely paramagnetic sectors. On the other hand, the $XX$ correlations have higher magnitudes but opposite signs in the sectors dominated by ferromagnetic and anti-ferromagnetic components. The nearest-neighbor concurrence behaves differently, with its largest values occurring where the transverse field and the Ising interaction compete. For the finite chains and parameter ranges checked numerically, we also find that the GGM is determined by a single-site bipartition. In these cases, the GGM increases as the magnitude of the transverse magnetization decreases. Finally, we have examined the sensitivity of the post-selected states to the Ising interaction strength through the quantum Fisher information. We find that additional peaks appear. This is due to the fact that different ground-state components approach their respective critical points, allowing the regions of enhanced sensitivity to be shifted. For the parameters and system sizes examined, the maximum quantum Fisher information is consistent with quadratic scaling in the system size, indicating Heisenberg scaling for the states being considered.

The present protocol provides a method to create superpositions of quantum many-body phases with the help of a central system. This protocol can be scaled up to higher dimensions for creating superpositions of many exotic quantum phases. These superposition states can be used in quantum sensing and information protocols that utilize multipartite entanglement. Moreover, the possibility of shifting the region of increased sensitivity by changing the coupling strength of the central spin provides an  opportunity to estimate parameters in the different interaction regimes. further studies could investigate if the coherence between quantum phases could give an advantage in quantum entanglement-based information protocols.
}
\acknowledgements
S.M. acknowledges support from the Infosys scholarship for senior students at Harish-Chandra Research Institute. U.S. acknowledges financial support from the Anusandhan National Research Foundation (ANRF), Government of India, under the Grant No. ANRF/ARG/2025/004617/PS.

\appendix
\section*{APPENDIX}



\section{Correlations and Magnetization}
\subsection{Nonzero correlations}
\label{AppB}
In this section, we explicitly calculate the various magnetizations and correlations discussed in the main text. We find that for superposition of Ising ground states, various observables are zero. This is discussed in App.~\ref{zero}. The only nonzero surviving correlations and magnetizations are \begin{equation}
\langle\sigma_i^z\rangle,\qquad
\langle\sigma_i^x\sigma_{i+1}^x\rangle,\qquad
\langle\sigma_i^y\sigma_{i+1}^y\rangle,\qquad
\langle\sigma_i^z\sigma_{i+1}^z\rangle.
\nonumber
\end{equation}

In order to calculate the various observables, we apply the Jordan-Wigner transformation~\cite{Mbeng2024}. Thus, for calculating the expecation values, we need to calculate the following,
\begin{equation}
\begin{aligned}
\langle a|
\sigma_j^x \sigma_{j+1}^x
|b\rangle
&=
\langle a|
\left(c_j^\dagger-c_j\right)
\left(c_{j+1}^\dagger+c_{j+1}\right)
|b\rangle,\\
\langle a|
\sigma_j^y \sigma_{j+1}^y
|b\rangle
&=
-\langle a|
\left(c_j^\dagger+c_j\right)
\left(c_{j+1}^\dagger-c_{j+1}\right)
|b\rangle,\\
\langle a|
\sigma_j^z \sigma_{j+1}^z
|b\rangle
&=
\langle a|
\left(1-2c_j^\dagger c_j\right)
\left(1-2c_{j+1}^\dagger c_{j+1}\right)
|b\rangle,\\
\langle a|
\sigma_j^z
|b\rangle
&=
\langle a|
\left(1-2c_j^\dagger c_j\right)
|b\rangle .
\end{aligned}
\end{equation}
Here $|a\rangle$ and $|b\rangle$ are the ground states of the quantum Ising model defined as 
\begin{equation*}
    |a\rangle = |g(J+a\delta)\rangle=\prod_{k>0}(u_k^{a}+iv_k^{a}c_k^{\dagger}c_{-k}^{\dagger})|0\rangle
\end{equation*}
\begin{equation*}
    |b\rangle = |g(J+b\delta)\rangle=\prod_{l>0}(u_l^{b}+iv_l^{b}c_l^{\dagger}c_{-l}^{\dagger})|0\rangle
\end{equation*}
where $u_k^{a}$, $v_k^{a}$ are defined as
\begin{equation}
    u_k^{a} = \frac{E_k^{a}+z_k^{a}}{\sqrt{2E_k^{a}(E_k^{a}+z_k^{a})}}
\end{equation}
\begin{equation}
    v_k^{a}=\frac{y_k^{a}}{\sqrt{2E_k^{a}(E_k^{a}+z_k^{a})}}
\end{equation}
where, $y_k^{a}=2J_a\sin{k}$, $z_k^{a}=2[h-J_a\cos{k}]$
and 
$E_{k}^{a}
=
 2\sqrt{
(h-J_{a}\cos k)^{2}
+
(J_{a}\sin k)^{2}
}$,
with, $J_a=(J+a\delta)$ and
\begin{equation}
    k=\pm\frac{(2n-1)\pi}{L} ,\quad \text{with} \quad n=1,...,\frac{L}{2}
\end{equation}

For calculating the $ZZ$ correlation, $C_1^{zz}$, we simplify to get 
\begin{align}
\langle a|
\sigma_j^z\sigma_{j+1}^z
|b\rangle
&=
\langle a|
\left(1-2c_j^\dagger c_j\right)
\left(1-2c_{j+1}^\dagger c_{j+1}\right)
|b\rangle \nonumber\\
&=
\langle a|b\rangle
-2\langle a|c_j^\dagger c_j|b\rangle
-2\langle a|c_{j+1}^\dagger c_{j+1}|b\rangle
\nonumber\\
&\quad
+4\langle a|
c_j^\dagger c_j
c_{j+1}^\dagger c_{j+1}
|b\rangle.
\end{align}
After using translation invariance of the model, we get
\begin{equation}
\langle a|
\sigma_j^z\sigma_{j+1}^z
|b\rangle
=
F^{ab}
-4\langle a|c_j^\dagger c_j|b\rangle
+4\langle a|
c_j^\dagger c_j
c_{j+1}^\dagger c_{j+1}
|b\rangle,
\label{ZZ-general}
\end{equation}
where, 
\begin{align}
F^{ab}=\langle a|b\rangle = \prod_{k>0} (u^a_ku^b_k + v^a_kv^b_k).
\label{eq:cross-fidelity}
\end{align}
\\
Then applying generalised Wick's theorem,
\begin{align}
\langle a|
c_j^\dagger c_j
c_{j+1}^\dagger c_{j+1}
|b\rangle
&=
\frac{
\langle a|c_j^\dagger c_j|b\rangle
\langle a|c_{j+1}^\dagger c_{j+1}|b\rangle
}{F^{ab}}
\nonumber\\
&\quad
-
\frac{
\langle a|c_j^\dagger c_{j+1}^\dagger|b\rangle
\langle a|c_j c_{j+1}|b\rangle
}{F^{ab}}
\nonumber\\
&\quad
+
\frac{
\langle a|c_j^\dagger c_{j+1}|b\rangle
\langle a|c_j c_{j+1}^\dagger|b\rangle
}{F^{ab}}.
\end{align}
Going to momentum space,
it can be shown that
\begin{equation}
\begin{aligned}
\langle a|c_j^\dagger c_{j+1}|b\rangle
&=
F^{ab}
\sum_{k>0}
\mathcal{X}_k^{ab}\cos k,\\
\langle a|c_j c_{j+1}^\dagger|b\rangle
&=
-{F^{ab}}
\sum_{k>0}
 \mathcal{X}_k^{ab} \cos k,\\
\langle a|c_j^\dagger c_{j+1}^\dagger|b\rangle
&=
-F^{ab}
\sum_{k>0}
 \mathcal{Y}_k^{ab} \sin k,\\
\langle a|c_j c_{j+1}|b\rangle
&=
F^{ab}
\sum_{k>0} \mathcal{Y}_k^{ba} \sin k,
\end{aligned}
\end{equation}
where
\begin{align}
\mathcal{X}^{ab}_k &= \frac{2}{L}\frac{v_k^a v_k^b}{M_k^{ab}},  \nonumber\\
\mathcal{Y}^{ab}_k &= \frac{2}{L}\frac{u_k^a v_k^b}{M_k^{ab}}, \nonumber \\
\mathcal{Y}^{ba}_k &= \frac{2}{L}\frac{u_k^b v_k^a}{M_k^{ab}},
\label{eq:X-mag}
\end{align}
with $M_k^{ab}=u_k^a u_k^b+v_k^a v_k^b$.
For the superscripts, we use $a,b \in\{-1,+1\}$ for spin-$1/2$ central system and $a,b \in \{-1,0,+1\}$ for spin-$1$ central system case. 
\\
\subsection{Vanishing correlations and expectation values}
\label{zero}
The transverse-field Ising Hamiltonian is invariant under the global $\mathbb{Z}_2$ parity transformation generated by
\begin{equation}
\hat{P}=\prod_{j=1}^{L}\sigma_j^z,
\nonumber
\end{equation}
under which the Pauli operators transform as
\begin{equation}
\hat{P}\sigma_j^x\hat{P}=-\sigma_j^x,\qquad
\hat{P}\sigma_j^y\hat{P}=-\sigma_j^y,\qquad
\hat{P}\sigma_j^z\hat{P}=\sigma_j^z.
\nonumber
\end{equation}
Since the ground state is an eigenstate of the parity operator, the expectation value of any parity-odd operator vanishes. Consequently,
\begin{equation}
\langle\sigma_j^x\rangle
=
\langle\sigma_j^y\rangle
=
\langle\sigma_j^x\sigma_{j+1}^z\rangle
=
\langle\sigma_j^z\sigma_{j+1}^x\rangle
=0.
\nonumber
\end{equation}
Furthermore, the Hamiltonian is real in the computational basis, and hence its ground state may be chosen to be real. Therefore, the expectation value of any operator containing an odd number of $\sigma^y$ operators also vanishes, yielding
\begin{equation}
\langle\sigma_j^x\sigma_{j+1}^y\rangle
=
\langle\sigma_j^y\sigma_{j+1}^x\rangle
=
\langle\sigma_j^y\sigma_{j+1}^z\rangle
=
\langle\sigma_j^z\sigma_{j+1}^y\rangle
=0.
\nonumber
\end{equation}
Hence, the only non-vanishing one- and two-spin expectation values are
$
\langle\sigma_j^z\rangle,
\langle\sigma_j^x\sigma_{j+1}^x\rangle,
\langle\sigma_j^y\sigma_{j+1}^y\rangle,
\langle\sigma_j^z\sigma_{j+1}^z\rangle.
$
\section{QFI}
\label{App-QFI}
In this section, we derive the QFI corresponding to the parameter $J$ encoded in a superposition of Ising ground state, given in Eq.~\eqref{eq:qfi} of the main text. For this, we
consider the state
\begin{equation}
|\psi_3(J)\rangle
=
\frac{1}{\sqrt{\mathcal{N}}}
\left[
c_{+1}|g(J+\delta)\rangle
+c_{0}|g(J)\rangle
+c_{-1}|g(J-\delta)\rangle
\right].
\end{equation}
This is generated using a central spin-$1$ system. The generated state has $J$ encoded. We also note that
the normalization $\mathcal{N}$ is also a function of the parameter $J$. It is given by
\begin{align}
\mathcal{N} = \sum_{a,b} c_a^* c_b F^{ab}, 
\end{align}
where $a,b\in \{+1, 0, -1\}$ with
\begin{equation}
F^{ab}=\langle g(J+a\delta)|g(J+b\delta)\rangle.
\end{equation}
We make the transformations $u_k^a = \cos(\theta^a_k/2)$ and $v_k^a = \sin(\theta^a_k/2)$. Thus, the overlap between two states is given by
\begin{equation}
F^{ab}
=
\prod_{k>0}
\cos\left(\Delta\theta_k^{ab}/2
\right).
\end{equation}
where $\Delta \theta_k^{ab} = \theta_k^{a} - \theta_k^{b}$.
\\
The quantum Fisher information with respect to $J$ is
\begin{equation}
\mathcal{F}_{Q}(J)
=
4\left[
\frac{\mathcal{A}}{\mathcal{N}}
-
\frac{|\mathcal{B}|^{2}}{\mathcal{N}^{2}}
\right],
\label{qfi}
\end{equation}
where
$\mathcal{A}
=
\langle\partial_{J}\psi|
\partial_{J}\psi\rangle$
and
$\mathcal{B}
=
\langle\psi|\partial_{J}\psi\rangle$.
The quantity $\mathcal{A}$ can be written as
\begin{align}
\mathcal{A} = \sum_{a,b} c_a^* c_b K^{ab}
\label{eq:A}
\end{align}
where
$K^{ab}
=
\langle\partial_{J}g(J+a\delta)
|
\partial_{J}g(J+b\delta)\rangle$.
The diagonal terms are
\begin{equation}
K^{aa}
=
4\sum_{k>0}
\frac{\sin^{2}k}
{(E_{k}^{a})^{4}}.
\end{equation}
For $a\neq b$, the cross terms are given by

\begin{align}
K^{ab}
&=
4F^{ab}
\Bigg[
\sum_{k>0}
\frac{\sin^{2}k \ \sec^{2}(
\Delta \theta_{k}^{ab}/{2})}
{(E_{k}^{a})^{2}(E_{k}^{b})^{2}}
\nonumber\\
&-\left(
\sum_{k>0}
\frac{\sin k \ \tan
({\Delta\theta_{k}^{ab}}/{2})}{(E_{k}^{a})^{2}}
\right)
\left(
\sum_{l>0}
\frac{\sin l \ \tan({\Delta\theta_{l}^{ab}}/{2})}{(E_{l}^{b})^{2}}
\right)
\Bigg],
\end{align}
Similarly, the quantity $\mathcal{B}$ is given by
\begin{align}
\mathcal{B} = \sum_{a,b}c_a^*c_b G^{ab},
\label{eq:B}
\end{align}
where
$G^{ab}
=
\langle g(J+a\delta)|\partial_{J}g(J+b\delta)\rangle$.
The diagonal terms are zero, i.e.,
\begin{equation}
G^{++}=G^{00}=G^{--}=0,
\end{equation}
while the cross terms with $a\neq b$ are given by
\begin{equation}
G^{ab}
=
2F^{ab}
\sum_{k>0}
\frac{1}
{(E_{k}^{b})^{2}}
\sin k \ \tan
({\Delta\theta_{k}^{ab}}/{2}).
\end{equation}

\bibliography{ref}

\end{document}